\documentclass[twocolumn,showpacs,preprintnumbers,amsmath,amssymb]{revtex4}

\usepackage{graphicx}   
\usepackage{dcolumn}   
\usepackage{bm}            
\usepackage{textcomp}
\usepackage{psfrag}
\usepackage{amsmath}
\usepackage{amsfonts}
\usepackage{amssymb} 
\usepackage[mathscr]{eucal}

{\rm }

\renewcommand{\vec}[1]{{\mathbf #1}}
\newcommand{\evec}[1]{{\hat{\mathbf #1}}}
\newcommand{\mat}[1]{{\mathcal #1}}

\renewcommand{\Re}[0]{\mathrm{Re}}
\renewcommand{\Im}[0]{\mathrm{Im}}

\begin{document}


\title{Traveling ion channel density waves affected by a conservation law}

\author{Ronny Peter}
\author{Walter Zimmermann}

\affiliation{
Theoretische Physik, Universit\"at Bayreuth, D-95440 Bayreuth, Germany
}

\date{\today}

\begin{abstract}
A model of mobile, charged ion channels embedded in a biomembrane is investigated. The ion channels fluctuate between an opened and a closed state according to a simple two--state reaction scheme whereas the total number of ion channels is a conserved quantity. Local transport mechanisms suggest that the ion channel densities are governed by electrodiffusion-like equations that have to be supplemented by a cable--type equation describing the dynamics of the transmembrane voltage. It is shown that the homogeneous distribution of ion channels may become unstable to either a stationary or an oscillatory instability. The nonlinear behavior immediately above threshold of an oscillatory bifurcation occuring at finite wave number is analyzed in terms of amplitude equations. Due to the conservation law imposed on ion channels large-scale modes couple to the finite wave number instability and have thus to be included in the asymptotic analysis near onset of pattern formation. A modified Ginzburg--Landau equation extended by long-wavelength stationary excitations is established and it is highlighted how the global conservation law affects the stability of traveling ion channel density waves.  
\end{abstract}

\maketitle

\section{\label{sec1} Introduction}
%
Spatiotemporal pattern formation is omnipresent in physical, chemical as well as biological systems driven out of equilibrium and can be be ascribed to a finite number of universality classes \cite{Manneville:1990,Cross:1993,Murray:1989,Cladis:1995,Zimmermann:1999,Weijer:1999,Weijer:2004}.
Cardiac spiral waves as well as the propagation of nerve signals along the axon of a neuron are prominent examples of bioelectric nonstationary patterns \cite{Glass:1981,Glass:1983,Gray:1995,Panfilov:1995,Echebarria:2002,Shiferaw:2005}; other examples might include the emergence of aster-like patterns in filament-motor-systems \cite{Nedelec:1997,Surrey:2001,Nedelec:2002,Ziebert:2005} or the nematic ordering of biopolymers such as actin and microtubule complexes \cite{Hitt:1990,Ziebert:2003}. Futhermore spontaneous pattern formation of ion channels is well-known to occur in cell membranes \cite{Jaffe:1977,Fromherz:1995,Leonetti:1997,Leonetti:2004} that are a ubiquitous building block in biology. In most biological systems the pattern forming processes are more elaborate than in fluid dynamical ones. Hitherto a qualitative, especially a quantitative understanding of universal aspects of pattern formation has mainly been achieved in fluid systems \cite{Cross:1993,Cladis:1995}. Biological systems, such as the model investigated in the present work, give however rise to new universal features of pattern formation. The global conservation law imposed upon the number of ion channels embedded into the cell membrane leads within the considered model system to a modified set of amplitude equations describing the generic properties of traveling ion channel density waves and alters moreover their stability behavior.

The considerable amount of information on membranes gained throughout the years has been synthesized into the fluid-mosaic model of a biomembrane \cite{Singer:1972} which includes the following main features: the phospholipids serve as a solvent for proteins and as a permeability regulator, the latter by adopting an energetically very effective bilayer configuration; membrane proteins are free to migrate within the lipid-bilayer; the conductance of the membrane layer is mainly made up of discrete ion channels formed by protein molecules \cite{Neher:1976}. Freely movable ion channels embedded in a fluid-mosaic membrane have been observed to self-organize if the concentration of salt across the membrane exceeds a certain threshold \cite{Fromherz:1988.1,Fromherz:1988.2,Fromherz:1991,Fromherz:1994}.

The basic fluid-mosaic model underwent however considerable refinement as it neglects first of all that ion channels may fluctuate between opened and closed states due to interactions with signal molecules \cite{Neher:1976} and secondly
that channel proteins may become immobilized due to binding to the cytoskeleton \cite{Peng:1986}. A compound system of a membrane with embedded ion channels is typically driven out of equilibrium due to ionic concentration gradients and transmembrane fluxes: Opened ion channels induce ionic transmembrane currents looped to the transmembrane voltage. Therefore thermodynamic models have to be extended by ion transport mechanisms based upon Nernst-Plank theory and by electrodynamical aspects \cite{Jaffe:1977,Leonetti:1997}. The resulting electrodiffusive models can be realized by considering a membrane seperating a narrow cleft of electrolyte from an electrolytic bath of ions \cite{Fromherz:1991.2}, a configuration often met in biological situations: closely neighboring cells like post-synaptic membranes of a neuronal synapse or membrane cables forming dendrites and axons of neurons are important examples of such geometries. The transmembrane voltage corresponding to the voltage difference across the thin layer of electrolyte is then given by a one- or two-dimensional cable-type equation \cite{Fromherz:1995,Leonetti:1998}.

It has been shown that the coupled dynamics of channel proteins and a binding-release reaction of the latter ones with the cytoskeleton leads to spatial peridoc patterns \cite{Fromherz:1995} described near onset of pattern formation by a Ginzburg-Landau equation \cite{Aranson:2002}. A model of two different, noninteracting species of ion channels has been studied in Refs.~\cite{Fromherz:1988.3,Hilt:2005}. A generic evolution equation for nonstationary patterns generated by two different ion channel species assumed to remain permanently in an opened state was given in Ref.~\cite{Hilt:2005} wherein a constant number of opened ion channels was presumed. In the present work we will consider 
ion channels that may switch between an opened and a closed state \cite{Hille:1992,Aidley:1996} according to a simple two--state reaction scheme with constant rates, the diffusion coefficients as well as the effective charges of the channel proteins being state--dependant. If compared to the models discussed in Refs.~\cite{Fromherz:1988.3,Hilt:2005}, the system at hand may be interpreted as one comprising two types of ion channels with a reaction stimulating transition between the different channel species while imposing the constraint of a conserved total number of ion channels.

After introducing the model in Sec.~\ref{sec2} the linear instability analysis performed in Sec.~\ref{sec3} evinces that either a stationary or an oscillatory bifurcation is met and is concluded by a phase diagram in Sec.~\ref{neutralcurves}. Immediately above the threshold of a supercritical bifurcation the amplitudes of the unstable pattern forming modes emerging from the homogeneous basic state are small enough to apply the powerful perturbational technique of amplitude equations. Oscillatory instabilities have been broadly discussed in literature as long as systems lack global conservation laws and their nonlinear evolution is governed by the most prominent class of amplitude equations, the so-called complex Ginzburg-Landau equations \cite{Cross:1993,Aranson:2002}. The influence of a conserved quantity on pattern forming instabilities, whether stationary or oscillatory, has been, starting from symmetry arguments, a subject of continuous interest during the recent years \cite{Hilt:2005,Cox:2000,Proctor:2001,Cox:2003,Cox:2004,Cox:2005} but was never discussed for a specific microscopic model. As stationary bifurcations have already been studied for the present biological system in Ref.~\cite{KramerSC:2002}, we will focus in the remaining sections on Hopf-bifurcations to traveling ion channel density waves. Within the context of a weakly nonlinear approach near onset of pattern formation in Sec.~\ref{sec4}, the complex Ginzburg-Landau equation known to describe stripe patterns is altered through the coupling to an evolution equation for a large-scale, real mode evoked by the presence of a global conservation law as shown in Sec.~\ref{genericamplitueeq}. In Sec.~\ref{nonlinearwaves} we investigate how the implication of the large-scale mode reflecting the conservation law modifies the stability regions of supercritically bifurcating traveling ion channel density waves obtained in Sec.~\ref{bifstructCGL}. It turns out that the latter ones become always unstable with respect to
an instability at finite wave number and that observing stable stripe patterns becomes less probable the more ion channels contribute to the total conductance of the membrane layer. Analytical calculations are confirmed by numerical simulations in Sec.~\ref{numerics}. Concluding remarks and an outlook on future work are given in Sec.~\ref{conclusion}.
%
\section{\label{sec2} Model system}
%
We consider a model system constituted of a membrane and embedded ion channels which may be encountered either in an opened or in a closed state. The dynamical equations of opened ($o$) and closed ($c$) ion channel densities $n_o({\bf r},t)$ respectively $n_c({\bf r},t)$ are coupled via a global conservation law $\int dS (n_o +n_c) = const.$ (wherein $dS$ denotes the area of the considered biomembrane) implemented by a simple local reaction scheme as well as to the cable equation governing the evolution of the transmembrane voltage.
%
\subsection{\label{sec2a} Basic equations}
%
The switch between the opened and closed state of an ion channel, referred to as opening--closing reaction, is governed by the following monomolecular reaction scheme 
\begin{align}
\begin{array}{lcr}
    & r_c                &     \\
n_c & \leftrightharpoons & n_o \\
    & r_o                &
\end{array}
\end{align}
with constant rates $r_c$ and $r_o$. According to the homogeneous, stationary ion channel densities
\begin{align}
\frac{\overline n_o}{\overline n_c}=\frac{r_o}{r_c}
\end{align}
channel dynamics are best captured in terms of the deviations $\tilde n_o$ and $\tilde n_c$ 
\begin{align}
\label{devia2}
\tilde n_o = n_o -\overline n_o\,, \qquad 
\tilde n_c = n_c -\overline n_c\,\,
\end{align}
from their respective homogeneous mean densities $\overline n_o$, $\overline n_c$. 

Local transport mechanisms as well as the mentioned conversion kinetics suggest the electrodiffusion-like equations of motion to be
\begin{subequations}
\label{devdyn}
\begin{align}
\partial_t  \tilde n_o \ &= - \nabla\vec j_o - r_c \tilde{n}_o + r_o \tilde{n}_c\, ,
\\ 
\partial_t  \tilde n_c \ &= - \nabla\vec j_c  + r_c \tilde{n}_o - r_o \tilde{n}_c
\, ,
\end{align}
\end{subequations}
wherein the current densities $\vec j_{o,c}$ depend on the chemical potential $\mu_{o,c}$ and the electrophoretic drift \cite{Jaffe:1977} via the transmembrane voltage $v$ as follows
\begin{align}
\label{current}
\vec j_{o,c}  &= - \nabla \mu_{o,c} - \nu_{o,c}\,q_{o,c}\,n_{o,c} \nabla v \, .
\end{align}
The occuring mobilities $\nu_{o,c}$ of opened and closed ion channels
\begin{align}
\label{mobilities}
\nu_{o,c}  &= \frac{D_{o,c}} {k_B T}\,
\end{align}
are proportional to the respective diffusion constants $D_{o,c}$. The stated charges $q_{o,c}$ of opened and closed ion channels are considered to be effective charges taking screening as well as electroosmotic effects into account \cite{Leonetti:1997}. If excluded volume interactions between opened and closed channels together with nonlinear effects are neglected, the chemical potential takes the simple form
\begin{align}
\label{potential}
\mu_{o,c}  &= D_{o,c}\,\nabla n_{o,c}\, .
\end{align}
Presuming a one--dimensional system substitution of the current $\vec j_{o,c}$ according to Eq.~(\ref{current}) in Eqs.~(\ref{devdyn}) gives rise to the following equations 
\begin{subequations}
\label{fundeq}
\begin{align}
\partial_t \tilde{n}_o &= D_o \nabla_x^2 \tilde{n}_o
+ \frac{D_o q_o}{k_B T} \nabla_x \left[(\tilde{n}_o + \bar{n}_o) \nabla_x v \right] -r_c \tilde{n}_o + r_o \tilde{n}_c\,,\\
\partial_t \tilde{n}_c &= D_c \nabla_x^2 \tilde{n}_c
+ \frac{D_c q_c}{k_B T} \nabla_x \left[(\tilde{n}_c + \bar{n}_c) \nabla_x v \right] -r_o \tilde{n}_c + r_c \tilde{n}_o
\end{align}
describing the spatiotemporal evolution of the ion channel densities. Those equations have nevertheless to be supplemented by an equation governing the transmembrane voltage $v$.

As a biomembrane is electrically equivalent to a cable whose repetitive basic unit is formed of a capacitor modeling the pecularities of lipid molecules, a voltage source reflecting the properties of an opened ion channel and a leak conductivity taking all remaining effects into account, the Kelvin cable equation may be considered to determine the transmembrane potential $v$ on condition that the caracteristic length scales of lateral patterns are large compared to the width of the electrolyte layer. A derivation of the one-dimensional cable equation 
\cite{Fromherz:1995} 
\begin{align}
C_m \partial_t v = \frac{1}{\Omega} \nabla_x^2 v - \Lambda_o n_o (v-E_o)
- G v\, ,
\end{align}
\end{subequations}
from the general Nernst-Plank theory was given in \cite{Leonetti:1998}. Herein $C_m$ denotes
the membrane capacitance, $\Omega$ the resistance of the thin membrane layer, $G$ the leak conductance and $\Lambda_o n_o$ the conductance of the opened ion channels per unit length. Finally $E_o$ stands for the Nernst potential, i.e. the resting potential of the membrane layer.

In most cases we consider the limit of a very quick response of the transmembrane voltage, i.e. formally $C_m=0$ holds, which is characteristic for physiological conditions \cite{Hille:1992}.
%
\subsection{\label{secscale} Scaled equations }
%
We rescale Eqs.~(\ref{fundeq}) by introducing dimensionless coordinates for space $x' = x / \overline{\lambda}$ and time $t'=t/\overline{\tau}$ with a typical length scale of an electrical perturbation  
$\overline{\lambda} = [\Omega (\Lambda_o {\overline n}_o +G)]^{-1/2}$ and the time constant of displacement $\overline{\tau} = \overline{\lambda}^2/D_o$. As normalized particle densities we opt for
\begin{align}
N_o := \frac{n_o - \bar{n}_o}{\overline{n}_o} = \frac{\tilde{n}_o}{\overline{n}_o}\,,
\qquad
N_c := \frac{n_c - \bar{n}_c}{\overline{n}_c} = \frac{\tilde{n}_c}{\overline{n}_c}\,,
\end{align}
the rescaled voltage reading $V := (v-v_R)q_o/k_BT$, wherein $v_R = \alpha E_o$
denotes the  resting voltage and 
$\alpha = \Lambda_o \overline{n}_o / (\Lambda_o \overline{n}_o +G ) \in\left[0,1\right]$
the density parameter measuring to what extent the ion channels contribute to the total conductance of the membrane layer. Through defining a normalized relaxation time $\tau_V = \Omega C_m D_o$, we obtain on the basis of the additional abbreviations
\begin{subequations}
\begin{align}
&\varepsilon = - \frac{q_o E_o}{k_BT},  \qquad
&\overline\tau = \frac{\overline{\lambda}^2}{D_o}\,,  \qquad
&D = \frac{D_c}{D_o}\,,\\
&R = \frac{q_c}{q_o}\,, \qquad
&\beta_o = \overline{\tau} r_o\,, \qquad
& \beta_c = \overline{\tau} r_c\,
\end{align}
\end{subequations}
the following scaled equations
\begin{subequations}
\label{scaleq}
\begin{eqnarray}
\label{scaleq1}
\partial_t N_o &=& \nabla_x^2 N_o + \nabla_x\left[(1+N_o)\nabla_x V\right]
\nonumber \\
&& \hspace{2cm} -\beta_c (N_o-N_c)\,, \\
\label{scaleq2}
\partial_t N_c &=& D\nabla_x^2 N_c + DR\nabla_x\left[(1+N_c)\nabla V\right]
\nonumber \\
&& \hspace{2cm} +\beta_o (N_o-N_c) \,,\\
\label{scaleq3}
\tau_V \partial_t V &=& (\nabla_x^2-1) V-\alpha(1-\alpha)\varepsilon N_o
- \alpha N_o V \label{scaleqcabl}
\end{eqnarray}
\end{subequations}
wherein the primes of the new coordinates $x'$ and $t'$ have been suppressed for the sake of simplicity.

It should be noticed that the spatially homogeneous basic state of these scaled equations is $N_o=N_c=V=0$.
%
\section{\label{sec3} Instability of the homogeneous state}
%
In a certain parameter range the spatially homogeneous basic state, namely $N_o=N_c=V=0$, becomes unstable against infinitesimal perturbations and a spatially inhomogeneous pattern is likely to emerge from this very state. In order to determine this interval of instability we transform the linear part of the partial differential equations Eqs.~\eqref{scaleq} by means of the ansatz
\begin{eqnarray}
\left( \begin{array}{ccc}
N_o \\ N_c \\ V
\end{array}\right) =
\left(\begin{array}{ccc}
\tilde N_o \\ \tilde N_c \\ \tilde V
\end{array} \right)
e^{\lambda t + i {\bf k} \cdot {\bf r}} \,
\end{eqnarray}
into a set of homogeneous linear equations for the amplitudes $\tilde N_o$, $\tilde N_c$
and $\tilde V$. These equations can be further simplified by assuming an instantaneously adapting transmembrane voltage $\tilde V$, i.e. presuming $\tau_V=0$ ($C_m\approx 1\frac{\mu F}{cm^2}$, $\Omega\approx 10^8\mathrm{Ohm}$, $D_o\approx 0.1\frac{\mu m^2}{s}$ \cite{Hille:1992}) or equivalently $C_m=0$. Within this limit the voltage perturbation $\tilde V$ may be expressed in the linear regime in terms of the opened ion channel density $\tilde N_o$
\begin{eqnarray}
\label{vlinear}
 \tilde V &=& -\,\dfrac{e}{1+k^2}\,\tilde N_o\, , 
\end{eqnarray}
with $e$ playing the role of the control parameter
\begin{align}\label{controlpara_e}
e = \alpha(1-\alpha)\,\varepsilon\, ,
\end{align}
and may thus be adiabatically eliminated from the other equations. For the two remaining linear and
homogeneous equations governing the pure ion channel dynamics
\begin{eqnarray}
\left(\begin{array}{cc}
\beta_c + k^2 -\dfrac{ek^2}{1+k^2} +\lambda &  -\beta_c \\ \\\hspace{-2mm}
-\beta_o -\dfrac{DRek^2}{1+k^2} & \hspace{-2mm}\beta_o +Dk^2 +\lambda
\end{array} \right)\left(\begin{array}{c}
\tilde N_o \\ \\ \tilde N_c  
\end{array} \right)=0\,,\nonumber \\
\end{eqnarray}
the solubility condition provides a quadradic polynomial for $\lambda$
\begin{eqnarray}
\label{quadpoly}
\lambda^2 + A \lambda + k^2 B =0\, ,
\end{eqnarray}
the two coefficients $A$ and $B$ reading
\begin{subequations}
\begin{eqnarray}
\label{defA}
A&=& \left[\beta_o+\beta_c + (1+D)k^2 - \dfrac{ek^2}{1+k^2} \right]\,\,,
 \\
\label{defB}
B &=& \beta_o +D\left( \beta_c +k^2 \right)
-e \,\dfrac{\beta_o + D(\beta_c R +k^2)}{1+k^2}\,.\nonumber \\
\end{eqnarray}
\end{subequations}
Depending on the model parameters the two solutions $\lambda_{1,2}$ of Eq.~\eqref{quadpoly} are either real
or complex. Provided that the real part of only one of those solutions $\lambda_{1,2}$ becomes positive in the neighborhood of a wave number $k$, the basic state $N_o=N_c=V=0$ is said to be unstable: In case of a vanishing imaginary part $\Im(\lambda)$ one has to deal with a {\it stationary} instability \cite{Cross:1993} whereas for finite imaginary parts, i.e. $\omega = \Im(\lambda)\not =0$, an oscillatory instability, a so--called {\it Hopf-bifurcation} \cite{Cross:1993}, is met.

Forthcoming discussions are considerably clarified by introducing the conversion rate ratio
\begin{align}
\label{convratio}
\beta=\frac{\beta_o}{\beta_c}=\frac{r_o}{r_c}\, .
\end{align}
%
%
\subsection{Dispersion relations}
\begin{figure}
\begin{center}
\includegraphics [width=0.97\columnwidth] {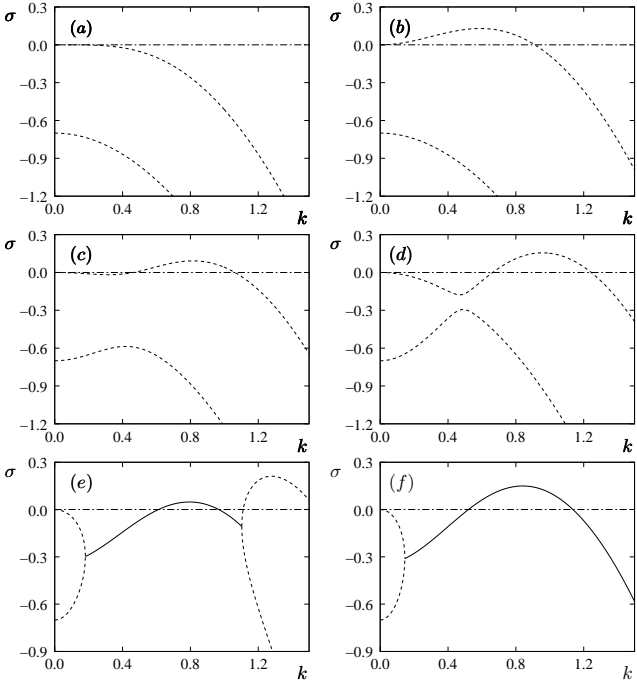}
\end{center}
\caption{\label{dispall}  The real parts of the two solutions
$\lambda_{1,2}(k)$ of Eq.~\eqref{quadpoly} are shown as a 
function of the wave number $k$ for different values of the 
charge ratio $R$ while the conversion rate ratio 
$\beta=\beta_o/\beta_c =0.4$ with $\beta_c =0.5$ and the diffusion
ratio $D=1$ are kept constant. Dashed branches correspond to 
stationary bifurcations, while solid ones to oscillatory 
instabilities. In part $(a)$ the charge ratio is $R=1.1$ and the 
control parameter has a subcritical value $e^S =0.93$,
whereas $e$ is always supercritical in the remaining cases.
In part $(b)$ the ratio $R=1.1$ and  $e^S =1.77$,
in $(c)$ $R=-0.1$ and $e^S =3.04$, in part $(d)$   $R=-0.39$ and
$e^S =3.69$, in part $(e)$ $R=-1.244$ and $e^H =5.32$
and finally in part $(f)$  $R=-1.8$ with $e^H =5.83$.
}
\end{figure}

%
The real parts $\sigma_{1,2}= \Re(\lambda_{1,2})$ of $\lambda_{1,2}(k)$, sketched in Fig.~\ref{dispall} for $D=1$ and $\beta=0.4$, exhibit typical shapes as a function of the wave number $k$. Beyond the threshold of pattern formation these different shapes of $\sigma_{1,2}(k)$ may lead in the weakly nonlinear regime to different universality classes of amplitude equations as discussed in the following paragraphs.

It should be noted that for typical Nernst potentials with absolute values reaching from $|E_o|=0$ to $|E_o|=100$ mV (equilibrium potentials  of the most important voltage gated ion channels for the mammalian skeletal muscle: $E_o=-98$ mV in case of $\mathrm{K}^{+}$ channels and $E_o=+67$ mV in case of $\mathrm{Na}^{+}$ channels \cite{Hille:1992}), the control parameter $e$ defined in Eq.~\eqref{controlpara_e} takes values in the range $|e|\in\left[0,10\right]$ under physilogical conditions. Thus all instabilities presented in Fig.~\ref{dispall} may play a role in a real biological system as electrophoretic charges are usually of a few elementary units.

Since the number of ion channels is a conserved quantity for the investigated model, i.e.
$\int dx (N_o+N_c)=const.$, the largest real part labelled $\sigma_{1}(k)$ is 
diffusive in the long-wavelength limit 
\begin{align}
\lambda_1 (k\to 0) = \sigma_1 (k\to 0) =- D_N k^2 + O(k^4)\, ,
\end{align}
with $D_N =\left[D\beta_c\left(Re-1\right)+\beta_o\left(e-1\right)\right]/\left(\beta_o+\beta_c\right)$ as confirmed by all examples shown in Fig.~\ref{dispall}.

In part $(a)$ of Fig.~\ref{dispall} the real parts of both eigenvalues, $\sigma_{1,2}(k)$, are drawn as a function of the wave number $k$ for a parameter  combination that leads to a stable homogeneous state. Part $(b)$ shows an example for the same charge and conversion rate ratio $R=1.1$ respectively $\beta=\beta_o/\beta_c=0.4$, with a positive curvature of the real part in reference to the wave number $k$
\begin{align}
\lambda_1 (k\to 0) = \sigma_1(k\to 0) = D_N k^2 \sim 0.9\,k^2 >0
\end{align}
and a maximum of the dispersion relation at a finite value $k_m\sim 0.6$. 
%
%
If the charge ratio $R$ is reduced while the conversion rate ratio $\beta$ remains
the same, the homogeneous state may become stable with respect to long-wavelength
perturbations as for the parameter set used in part $(c)$, whereas the curvature
$\left.\partial_k \sigma_1(k)\right|_{k=0} <0$ is negative. The homogeneous state gets
nevertheless unstable against perturbations with a finite wave number in the 
vicinity of $k=k_c=0.8$ as can be foreseen from the extremum condition 
$\left.\partial_k \sigma_1(k)\right|_{k=k_c} = 0$.

%
%
As the charge ratio $R$ is further diminished, the two real parts $\sigma_{1,2}(k)$
approach each other as depicted in part $(d)$. Decreasing $R$ even further leads to a
pair of complex eigenvalues $\lambda_{1,2}$ instead of real ones [cf. Fig.~\ref{dispall} $(e)$],
meaning that the homogeneous basic state becomes simultaneously unstable against a pair
of traveling waves (solid branch) with a critical wave number of $k\sim 0.8$ and a spatially periodic modulation (dashed branch) with a wave number of $k \sim 1.35 $.
As $R$ gets even smaller the latter stationary branch, i.e. the dashed one visible in part $(e)$, becomes strongly negative and only a Hopf--bifurcation to traveling waves remains as depicted in part $(f)$. When one of the dispersion curves in Fig.~\ref{dispall} hits the $k-$axis as a function of the control parameter $e$, the threshold of the respective instability is met.
%
%
%
\subsection{Stationary threshold}
%
The threshold $e_c^S$ for a stationary bifurcation follows from the neutral stability condition $\sigma=\lambda=0$, which is equivalent to claiming $B=0$ in Eq.~\eqref{quadpoly}. From  $B=0$ the wave number dependent neutral curve \cite{Cross:1993, Manneville:1990}
\begin{equation}
\label{eS}
e^S(k) = \dfrac{\left[D\left(k^2 +\beta_c\right) +\beta_o\right]\left(1+k^2\right)}
            {D\left(k^2 +R\beta_c\right) +\beta_o}
\end{equation}
may be derived. This neutral curve takes its minimum either at a vanishing critical wave number $k_c^S=0$
\begin{equation}
\label{eSc0}
e^S(0) = \dfrac{D\beta_c +\beta_o}
            {DR\beta_c+\beta_o}
\end{equation}
or the minimum becomes even lower than the latter one for a finite value $k_c^S$ of the wave number
\begin{equation}
\left(k_c^S\right)^2 = -R\beta_c -\dfrac{\beta_o}{D}
              +\dfrac{\sqrt{\beta_c\left(R-1\right)\left[D\left(R\beta_c -1\right)+\beta_o\right]}}
              {\sqrt{D}} \,,
\end{equation}
where $k_c^S$ is given by the extremum condition $\partial_k e^S(k)=0$. The threshold value of the control parameter $e$ at $k_c^S$ reads 
\begin{equation}\label{eSc}
e^S_c =\dfrac{\left[D\beta_c\left(R-1\right)-\Gamma
              \right]
              \left
[D\left(R\beta_c -1\right)+\beta_o-\Gamma
              \right]}
             {D\Gamma}
\end{equation}
wherein  the abbreviation
\begin{equation}
\Gamma = \sqrt{D\beta_c\left(R-1\right)\left[D\left(R\beta_c -1\right)+\beta_o\right]}
\end{equation}
has been introduced.
%
\subsection{Hopf-bifurcation}

\begin{figure}
\includegraphics [width=0.9\columnwidth] {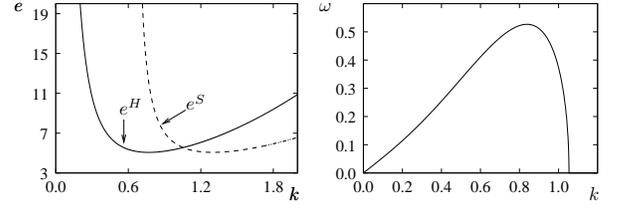}
\caption{\label{hopfstatct}
Neutral curves for a stationary ($e^S$) and oscillatory instability ($e^H$) as a function of the wave number $k$ (on the left hand side) and the frequency along these neutral curves 
(on the right hand side) as given by Eq.~\eqref{wH}
for the parameter set used in Fig.~\ref{dispall}$(e)$.
}
\end{figure}

At onset of a Hopf--bifurcation the real parts $\sigma_{1,2}$ of the complex conjugate pair of eigenvalues $\lambda_{1,2}$ vanish, while the imaginary parts 
$\omega_{1,2} = \pm \Im(\lambda)$ remain nevertheless finite. Assuming complex eigenvalues $\lambda_{1,2}$ the polynomial in Eq.~\eqref{quadpoly} may be decomposed into its real and imaginary part. Since both parts have to vanish separately in order to satisfy the solubility condition Eq.~\eqref{quadpoly}, they provide two conditions allowing to determine the neutral curve of the oscillatory instability
\begin{equation}\label{eH}
e^H(k) = \frac{\left(1+ k^2\right)\left(k^2 + D k^2 + \beta_o + \beta_c\right)}
            {k^2}\, ,
\end{equation}
as well as the frequency
\begin{equation}\label{wH}
\omega^2(k) = k^2\,B(k,e)
\end{equation}
with $B$ from Eq.~\eqref{defB}. Both quantities are shown as a function of the wave number $k$ for a typical
parameter combination in Fig.~\ref{hopfstatct} where the stationary and the Hopf-branch have a similar
threshold as expected from Fig.~\ref{dispall}$(e)$. If parameters corresponding to Fig.~\ref{dispall}$(f)$ are chosen, the shape of the neutral curves depicted in Fig.~\ref{hopfstatct} would not be altered but the Hopf-branch would have the lowest threshold.
The neutral curve specified in Eq.~\eqref{eH} takes its minimum at 
\begin{equation}\label{kHc}
k_c^H =   \left(\dfrac{\beta_o +\beta_c}{1 + D}\right)^{\dfrac{1}{4}}
\end{equation}
and the corresponding critical value of the control parameter at $k_c^H$ is
\begin{equation}\label{eHc}
e_c^H = \left(\sqrt{1+D} +\sqrt{\beta_o +\beta_c}\right)^2 \, ,
\end{equation}
whereas the Hopf--frequency evaluates to
\begin{align}
\omega_c=k_c^H\sqrt{\gamma\sqrt{1+D}-D{k_c^H}^2(D+2\beta_o)}
\end{align}
with
\begin{align}
\gamma=\beta_o\sqrt{\beta_o+\beta_c}+R\beta_c {e_c^H}^2\, .
\end{align}
%
%
%
\subsection{Neutral curves and phase diagram}
\label{neutralcurves}
\begin{figure}
\includegraphics [width=0.95\columnwidth] {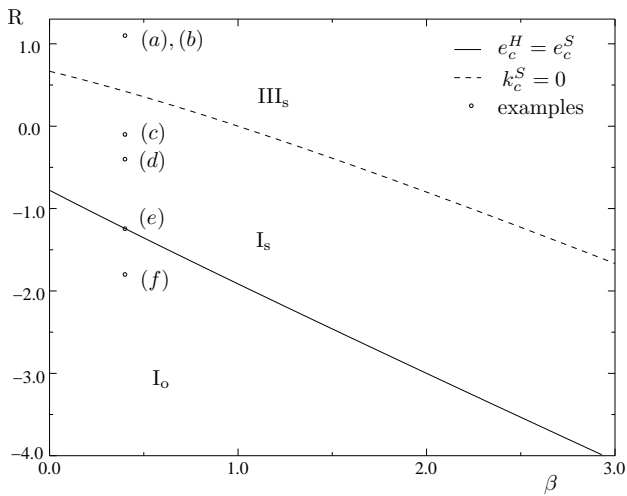}
\caption{\label{phasedia1} The figure shows the respective ranges
of a stationary bifurcation either of the Cahn--Hilliard-type with a critical
wave number $k_c=0, \omega_c=0$ or with a finite critical wave number
$k_c \not =0, \omega_c=0$ and of a Hopf-bifurcation $k_c\not=0, \omega_c\not=0$.
The symbols $(a)$ to $(f)$ correspond to the  parameters in the $R$-$\beta$-plane 
for whom the dispersion relations are plotted in Fig.~\ref{dispall}, parameters
being $D=1$ and $\beta_c=0.5$. 
}
\end{figure}

%
There are obviously three different types of instabilities of the homogeneous state occuring in different areas of the $R$-$\beta$-plane, $\beta$ being the conversion rate ratio defined in Eq.~\eqref{convratio}. One has either a Cahn--Hilliard-like \cite{Cahn:1957} instability characterized by $e^S(k=0)< e^S_c$ and $e^S(k=0)< e^H_c$ ($\mathrm{III_s}$-bifurcation \cite{Cross:1993}),
i.e. the long-wavelength excitations dominate; a stationary instability at a finite 
wave number $k$ with $e^S_c < e^S(k=0)$ as well as $e^S_c <e^H_c$ ($\mathrm{I_s}$-bifurcation \cite{Cross:1993}); or a Hopf--bifurcation distinguished by $e^H_c < e^S(k=0)$ and $e^H_c < e^S_c$ ($\mathrm{I_o}$-bifurcation \cite{Cross:1993} ). These three regions of existence are shown in Fig.~\ref{phasedia1} for $\beta_c=0.5$ and $D=1$, wherein the dashed line is determined by the condition that the two stationary thresholds agree
\begin{equation}
\label{ssboundary}
e^S_c = e^S(k=0) \, ,
\end{equation}
whereas the solid line is given by the codimension-2 condition
\begin{equation}
\label{ssboundary}
e_c^H = e_c^S\,.
\end{equation}
The dispersion relation corresponding to the point labelled $(c)$ in Fig.~\ref{phasedia1} is given by Fig.~\ref{dispall}$(c)$. The parameter set corresponding to $(e)$ in Fig.~\ref{phasedia1} lies in the 
immediate vicinity of the codimension-2 line and the neutral curves as well as the change of frequency along those curves are drawn as a function of the wave number $k$ in Fig.~\ref{hopfstatct}, the associated dispersion relation being depicted in Fig.~\ref{dispall}$(e)$.

\begin{figure}
\includegraphics [width=0.95\columnwidth] {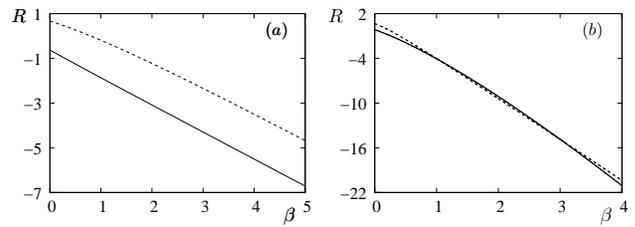}
\caption{\label{phasedia2} The boundary between a $\mathrm{I_s}$-bifurcation
and a $\mathrm{III_s}$-bifurcation (dashed line) as well as the one corresponding to
the transition from a stationary to an oscillatory bifurcation (solid line) are shown for
different diffusion ratios $D=0.8$ in (a) and $D=0.18$ in (b) while the conversion rate $\beta_c=0.5$.
}
\end{figure}

If the diffusion ratio $D$ is reduced compared to $D=1$ in Fig.~\ref{phasedia1}, while the conversion rate $\beta_c$ is nevertheless kept constant at $\beta_c=0.5$, the area wherein a $\mathrm{I_s}$ stationary bifurcation occurs is mostly affected as depicted in Fig.~\ref{phasedia2}. For large values of the diffusion ratio, such as $D=0.8$ in Fig.~\ref{phasedia2}(a), a stationary bifurcation at finite wave number is encountered in a broad band of the $R$-$\beta$-plane bounded by the dashed line corresponding to the border between the $\mathrm{III_s}$- and $\mathrm{I_s}$-bifurcation and the solid line visualizing the codimension-2 line. Reducing $D$ induces the region of existence of the $\mathrm{I_s}$-bifurcation to shrink significantly. For very small diffusion ratios, such as $D=0.18$ in Fig.~\ref{phasedia2}(b), a broadening of the $\mathrm{I_s}$-bifurcation-region occurs for large values of $\beta$ or equivalently for strongly negative values of $R$, which is not so pronounced for intermediate diffusion ratios $D$, while its area of existence remains strongly confined for slightly negative charge ratios. For $1\lesssim\beta\lesssim 3$ the solid line seperates in case of $D=0.18$ [Fig.~\ref{phasedia2}(b)] a Hopf-bifurcation from a $\mathrm{III_s}$-bifurcation instead of a $\mathrm{I_s}$-bifurcation otherwise, such an interval of $\beta$ values can always been determined as long as $D\lesssim 0.3$. 

Furthermore it should be noticed from Fig.~\ref{phasedia2} that with decreasing values of $D$ the slopes of the codimension-2 line and of the phase limit between the $\mathrm{III_s}$- and $\mathrm{I_s}$-bifurcation become steeper which progressively shifts the domain of emergence of an oscillatory bifurcation, lying beneath the codimension-2 line, to more negative charge ratios $R$.

\begin{figure}
\includegraphics [width=0.95\columnwidth] {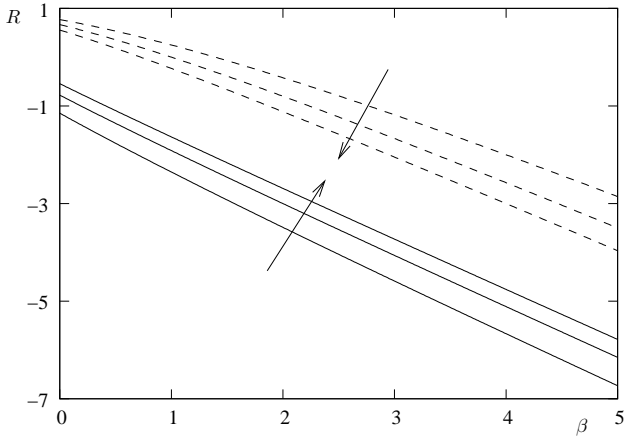}
\caption{\label{phasedia3} Phase diagram in the $R$-$\beta$-plane for different values of the conversion rate $\beta_c$ given by Eq.~(\ref{convratio}). The codimension-2 lines (solid lines) as well as the phase boundaries between a $\mathrm{I_s}$-bifurcation and a $\mathrm{III_s}$-bifurcation (dashed lines) are displayed for  $\beta_c=0.3$, $\beta_c=0.5$ and $\beta_c=0.8$. The directed arrows visualize the shrinking of the area wherein a $\mathrm{I_s}$-bifurcation is encountered with growing values of the conversion rate $\beta_c$. The diffusion
ratio $D$ is assumed to evaluate to 1.
}
\end{figure}

If the diffusion ratio $D$ is however kept constant while the conversion rate $\beta_c$ is varied, the relative positions of the phase boundaries remain almost unchanged as it may be observed in Fig.~\ref{phasedia3} wherein $D=1$. The latter figure evinces that the area of occurence of a $\mathrm{I_s}$ stationary bifurcation located in--between the two phase boundaries gets insignificantly smaller for increasing values of $\beta_c$. Hence the value given to the conversion rate $\beta_c$ is not determinant while focusing on Hopf--bifurcations in the forthcoming sections.
%
\section{Weakly nonlinear analysis beyond a Hopf-bifurcation}
\label{sec4}
%
We will focus in the present section on the oscillatory bifurcation regime while a partial analysis of the 
$\mathrm{I_s}$-bifurcation may be found in Ref.~\cite{KramerSC:2002}. Immediately beyond the threshold of a supercritical bifurcation, where the amplitude of a pattern emerging from the homogeneous state is still small, the concept of the so-called amplitude equations is a very successful one to characterize the nonlinear behavior of patterns as it is exemplified for several physical, chemical and biological systems in Refs. \cite{Cross:1993, Manneville:1990}.

Due to the presence of the global conservation law, namely $\int\,dx\,(N_o+N_c)=const.$, all dispersion relations depicted in  Fig.~\ref{dispall} behave diffusively in the long-wavelength limit. Therefore large-scale modes are 
marginal at the considered bifurcation point and have to be included in the reduced dynamics near onset of pattern formation. Restricting our investigation to traveling ion channel density waves, the complex Ginzburg--Landau equation \cite{Aranson:2002} normally used to describe traveling waves has to be extended by an evolution equation for the large-scale mode that is evoked by the aforementioned global conservation law. This set of coupled amplitude equations, is obtained in Sec.~\ref{genericamplitueeq} by a systematic perturbation expansion from the scaled equations Eqs.~\eqref{scaleq}, whereby the coefficients occuring in these amplitude equations reflect the dependance of the pattern on the parameters of the underlying system. The symmetry consitstency of the latter equations is checked in Sec.~\ref{genericamplitueeq} as well.

After having studied the bifurcation structure by means of the complex Ginzburg--Landau equation in Sec.~\ref{bifstructCGL}, it will be shown in Sec.~\ref{nonlinearwaves} that the implication of the large-scale mode dramatically changes, in case of an infinitely extended system, the results on stability of traveling waves acquired in Sec.~\ref{bifstructCGL}. Finally the validity of the performed amplitude expansion is checked numerically in Sec.~\ref{numerics}.
%
\subsection{Generic amplitude equations}
\label{genericamplitueeq}
%
%
The small parameter used in the perturbative derivation of the coupled amplitude equations is the relative distance $\eta$ to the threshold $e_c$ or likewise $\varepsilon_c$
\begin{equation}
\eta =\dfrac{e-e_c}{e_c}=\dfrac{\varepsilon-  \varepsilon_c}{\varepsilon_c}\, .
\end{equation}
Given a system presenting a global conservation law, as encountered in this work, the generating field $\vec{w}_1=(N_o, N_c)^T$ of traveling waves  in the vicinity of the bifurcation point comprises a complex, propagating mode $A$ and a real, large-scale mode $C$ stipulated by the constraint of conserved number of ion channels
\begin{eqnarray}
\label{amplinlo}
\vec{w}_1 = A\,\evec{e}_0\,
e^{i(\omega_c  t + {\bf k_c^H} \cdot\vec r)}  \,
+ C\,\evec{v}_0 + c.c.\, ,
\end{eqnarray}
wherein $c.c.$ denotes the complex conjugate of the preceding terms. Comparable to the adiabatic elimination of the transmembrane voltage $V$ performed via Eq.~(\ref{vlinear}) in Sec.~\ref{sec3} within the linear regime, $V$ can be similarly expressed in terms of the channel densities within the weakly nonlinear regime, the field $\vec{w}_1$ becoming thus implicitly dependant of the membrane voltage. $\vec{k_c^H}$ in the latter ansatz designates the critical wave number defined in Eq.~\eqref{kHc} and the cited eigenvectors are
\begin{align}\label{eigenvectors}
\evec{e}_0 = \left(\begin{array}{c} 1\\F_0      \end{array}\right)\, ,\qquad
\evec{v}_0 =\left(\begin{array}{c} 1\\1\end{array}\right)\, ,
\end{align}
the second component of the eigenvector $\evec{e}_0$ reading
\begin{equation}
F_0 = \dfrac{-D\sqrt{\beta_o+\beta_c}-\beta_o\sqrt{1+D}+\dot{\imath}\omega_c\sqrt{1+D}}
            {\beta_c\sqrt{1+D}}\, .
\end{equation} 
The amplitudes $A$ and $C$ of the fluctuations $\vec{w}_1$ depend on the following critically slow time scales $T_1=\sqrt{\eta}\,t$ and $T=\eta\,t$ as well as on the scaled spatial variable $X=\sqrt{\eta}\,x$, thus suggesting a standard multiscale perturbative approach \cite{Cross:1993}, that has to be combined with the aforesaid adiabatic elimination of the transmembrane voltage as sketched for convinience in Appendix~\ref{appA}.

The expansion of the fundamental equations Eqs.~\eqref{scaleq} and of the pattern forming fields $N_o$, $N_c$ as well as $V$ with respect to powers of the small parameter $\sqrt{\eta}$ results on the basis of the ansatz specified in Eq.~\eqref{amplinlo} in the subsequent system of coupled amplitude equations
\begin{subequations} \label{amplitudeeq}
\begin{align}
\tau\left(\partial_t +v_g \partial_x \right)A
&=\left(\eta +\dot{\imath} a\right) A
+\xi_0^2\left(1+\dot{\imath} c_0\right)\partial_x^2 A
\nonumber\\
&\hspace{0.2cm}-g_1\left(1+\dot{\imath} c_1\right) A|A|^2-g_2\left(1+\dot{\imath} c_2\right) A\,C
\nonumber\\
&\hspace{0.2cm}- g_3\left(1+\dot{\imath} c_3\right)C^2 A+g_4\left(1+\dot{\imath} c_4\right) C\partial_x A
\nonumber\\
&\hspace{0.2cm}+g_5\left(1+\dot{\imath} c_5\right) A\partial_x C\, ,\label{amplitudeeq_1}\\
\partial_t C
&= D_1 \partial_x^2 C - b\, \partial_x |A|^2\, , \label{amplitudeeq_2}
\end{align}
\end{subequations}
provided that space and time are once more rescaled to $x$ and $t$. 

\begin{figure}
\includegraphics [width=0.99\columnwidth] {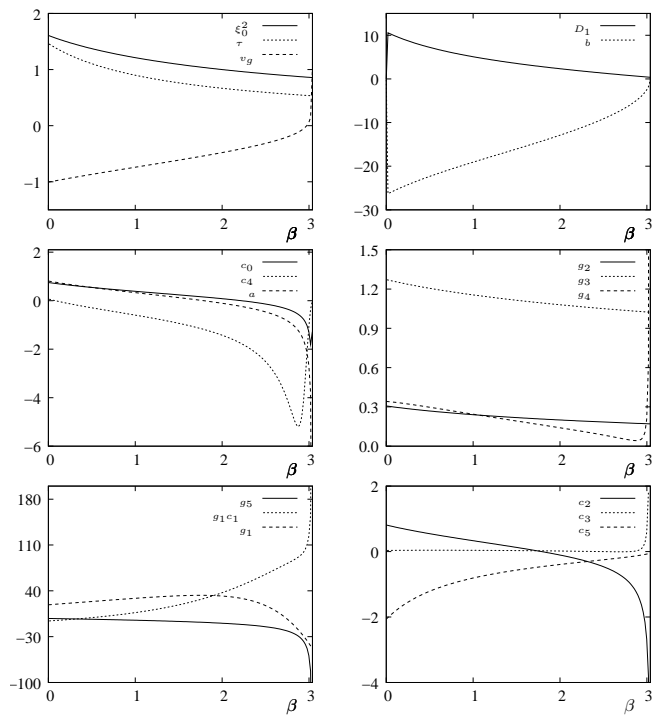}
\caption{\label{coeffplots}
The coefficients of the coupled amplitude equations Eqs.~\eqref{amplitudeeq} are plotted as functions of the conversion rate ratio $\beta$ for the parameter combination $D=0.5$, 
$\alpha=0.8$, $R=-5.5$ and $\beta_c=0.5$.
}
\end{figure}

The analytical expressions of the coefficients appearing in the latter equations Eqs.~\eqref{amplitudeeq} as functions of the parameters of the underlying biological model are rather lengthy. Instead of giving their analytical forms we therefore plot them in Fig.~\ref{coeffplots} as functions of the conversion rate ratio $\beta$ while keeping the remaining parameters constant.

Likewise amplitude equations, wherein the real, however damped, mode $C$ plays the role of a concentration field, were used to describe localized traveling-wave trains observed in experiments on binary mixture convection \cite{Riecke:1992} and to clarify the influence of a long-wave mode on solitary waves \cite{Riecke:1996}. The amplitude equations discussed in the latter references were recovered by a reaction-diffusion system \cite{Ipsen:2000}. In Ref. \cite{Riecke:2001} the nonlinear behavior of traveling waves arising in a supercritical Hopf-bifurcation, coupled to a real, weakly damped mode $C$ that is advected by the waves and thus affects the stability of the latter ones, yielded envelope equations resembling to Eqs.~(\ref{amplitudeeq}). In all the previous examples the real mode $C$ was nevertheless damped and not a true zero mode, i.e. a conserved quantity: Slightly different amplitude equations, motivated by work on magnetoconvection as well as rotating convection, with a true conserved quantity were derived by symmetry arguments in Ref.~\cite{Cox:2005}. The amplitude equations of Ref.~\cite{Cox:2005} as well as those found for the present biological model will be further discussed within the scope of a two-component reaction-diffusion model 
which also describes the coupling of a traveling wave to an oscillatory long-wavelength mode instead of a stationary one and leads thus to a generalized version of Eqs.~(\ref{amplitudeeq}) \cite{Peter:2006}.
%
%
%

Let us now check if the amplitude equations Eqs.~(\ref{amplitudeeq}) obtained by the multiscale perturbative approach are consistent with the symmetries of the underlying model system. The electrodiffusive model equations (\ref{fundeq}) or equivalently the scaled basic equations (\ref{scaleq}) present translational as well as reflectional symmetry. Accordingly amplitude equations describing standing waves generated by two complex counterpropagating linear modes $A_1(x,t)$ and $A_2(x,t)$ coupled to a real, large-scale field $C(x,t)$ evoked by the conservation law inherit the following invariance under spatial and temporal translation:
\begin{subequations}
\label{transsym}
\begin{align}
x\mapsto x+\delta: 
& 
& A_1\rightarrow A_1 e^{\dot{\imath}\theta},\,
& A_2 \rightarrow A_2 e^{-\dot{\imath}\theta}, 
& C \rightarrow C,\label{transsym1}\\
t\mapsto t+\kappa:
&
& A_1\rightarrow A_1 e^{\dot{\imath}\phi},\,
& A_2 \rightarrow A_2 e^{\dot{\imath}\phi},
& C \rightarrow C,\label{transsym2}
\end{align}
\end{subequations}
where $\theta=k_c^H\delta$, $\phi=\omega_c\kappa$ as well as under reflection:
\begin{align}\label{refsym}
x\mapsto -x: A_1\rightarrow A_2,\,\, A_2\rightarrow A_1,\,\, C\rightarrow C\, .
\end{align}
Due to the scalar constraint of a total conserved number of ion channels the translational symmetry, c.f. Eq.~(\ref{transsym1}), stipulates that the leading-order coupling term that extends the complex Ginzburg--Landau equations \cite{Aranson:2002,Cross:1993}
\begin{align}
\tau\left(\partial_t \pm v_g \partial_x \right)A_i
=& \left(\eta +\dot{\imath} a\right) A_i +\xi_0^2\left(1+\dot{\imath} c_0\right)\partial_x^2 A_i
\nonumber\\
&\hspace{0.05cm}-g_1\left(1+\dot{\imath} c_1\right) A_i\,|A_i|^2\nonumber\\
&\hspace{0.05cm}-g_6\left(1+\dot{\imath} c_6\right) A_i\,|A_j|^2\, ,\nonumber\\
&\hspace{0.6cm}\mathrm{with}\,\,i,j\in\left\lbrace 1,2\right\rbrace\,\mathrm{and}\,j\neq i\,\, ,
\end{align} 
is proportional to $A_i\,C$ (i=1,2). Possible higher order terms that might be included in extended Ginzburg--Landau equations comprise $A_i\,C^2$ or $C(\partial_x A_i)+A_i\partial_x C$ with $i=1,2$. In order to satisfy the reflectional symmetry given by Eq.~\eqref{refsym} the complex coefficents of the latter two terms need to be antisymmetric with respect to reflections as they involve first order spatial derivatives.

In absence of eventual coupling terms the real mode $C$ fulfills a diffusion equation as suggested by the conserved quantity and the reflectional symmetry specified in Eq.~(\ref{refsym}). The translational symmetries cited in Eqs.~(\ref{transsym}) then claim that coupling terms from the propagating modes $A_1$ and $A_2$ likely to occur in the evolution equation of $C$ involve equal numbers of $A_i$ and their associated complex conjugates $A_i^{\star}\,\,(i=1,2)$. Therefore the leading coupling term consistent with reflectional symmetry reads
$\partial_x(|A_1|^2-|A_2|^2)$, the generic evolution equation for $C$ finally being a diffusion equation with nonlinear forcing from $A_i\,(i=1,2)$.

Within the limit $A_1\rightarrow A$ and $A_2\rightarrow 0$, the standing wave patterns reduce to traveling waves and the symmetry consistent amplitude equations happen to be exactly those, namley Eqs.~\eqref{amplitudeeq}, obtained by a standard multiscale approach from the inital model equations Eqs.~\eqref{scaleq} in the previous section. Furthermore the coefficients $g_4$, $g_5$ and $b$ appearing in Eqs.~\eqref{amplitudeeq} are pseudoscalars as required by the reflectional symmetry of the underlying model system.
%
\subsection{Bifurcation structure of traveling waves
in absence of the large-scale mode $C$}
\label{bifstructCGL}
%
Since the large scale mode $C$ only affects stability properties of traveling waves, we may neglect $C$ while determining their general bifurcation structure.
The problem at hand then corresponds to a simple traveling wave of complex amplitude $A$ and group velocity $v_g$ described by the amplitude equation
\begin{align}\label{amplitravwave}
\tau\left(\partial_t +v_g \partial_x \right)A
=& \left(\eta +\dot{\imath} a\right) A +\xi_0^2\left(1+\dot{\imath} c_0\right)\partial_x^2 A
\nonumber\\
&\hspace{0.2cm}-g_1\left(1+\dot{\imath} c_1\right) A|A|^2\, . 
\end{align}
%
By changing to a coordinate frame moving with the group velocity $v_g$, i.e. by performing a Gallilei transform, 
Eq.~\eqref{amplitravwave} can be reduced to the prominent complex Ginzburg--Landau equation \cite{Aranson:2002,Cross:1993}
\begin{align}\label{ginzlandau}
\partial_t A
= \left(\eta +\dot{\imath} a\right)A+\left(1+\dot{\imath}c_0\right)\partial_x^2 A
-g_1\left(1+\dot{\imath} c_1\right)A|A|^2\, ,
\end{align}
wherein the linear imaginary part $\dot{\imath} a$ is easily eliminated by the transformation
$A\rightarrow A\,e^{\dot{\imath} a t}$.
%
\subsubsection{Supercritical vs. subcritical bifurcation}
%
For the complex Ginzburg--Landau equation it is well known that traveling waves bifurcate supercritically if the real part of the coefficient of the cubic term $A|A|^2$ in Eq.~\eqref{ginzlandau} is positive, namely $g_1>0$.
The tricritical line, along which the coefficient $g_1$ vanishes, marks the border between super- and subcritically bifurcating waves in parameter space \cite{Aranson:2002}.

\begin{figure}
\includegraphics [width=0.85\columnwidth] {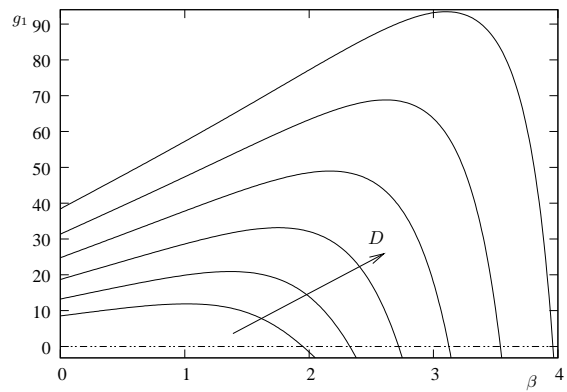}
\caption{\label{tricrit2} The coefficient $g_1$ appearing in the amplitude
equations Eqs.~\eqref{amplitudeeq} is drawn as a function of the conversion
rate ratio $\beta$ for different diffusion ratios $D$, with 
$D=0.3,0.4,0.5,0.6,0.7,0.8$ from left to right. The zero points illustrate
how the tricitical point is moving for a given charge ratio $R=-5.5$ to increasing values 
of $\beta$. Further parameters are $\beta_c=0.5$, $\alpha=0.8$. 
}
\end{figure}

\begin{figure}
\includegraphics [width=0.85\columnwidth] {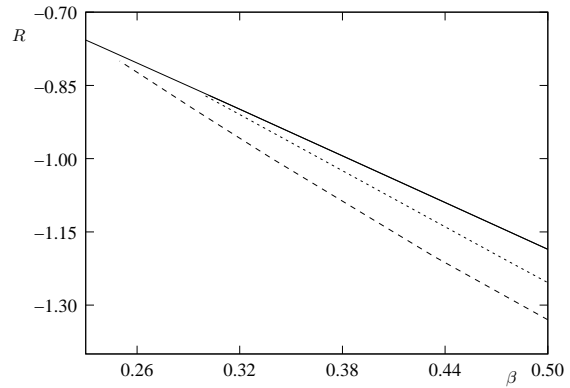}
\caption{\label{tricrit1} The codimension-2 line (solid line) as well
as the tricritical lines are plotted in the $R$-$\beta$-plane for the
density parameters $\alpha=0.2$ (dashed line) and $\alpha=0.8$ (dotted line).
Furthermore $D=0.5$ and $\beta_c=0.5$.
}
\end{figure}

Fig.~\ref{tricrit2} shows how the tricritical point, namely the root of the coefficient $g_1$, moves at a strongly negative charge ratio $R=-5.5$ towards greater values of the conversion rate ratio $\beta$ as $D$ becomes larger.
For a given value of the diffusion ratio, such as $D=0.5$, and a fixed conversion rate $\beta_c=0.5$ the tricritical line is only slightly shifted within the $R$-$\beta$-plane while varying the density parameter $\alpha$ over the permitted interval ($\alpha\in[0,1]$) as may be concluded from Fig.~\ref{tricrit1}. Traveling waves always bifurcate supercritically beneath the tricritical line (dashed and dotted lines in Fig.~\ref{tricrit1}) and a change to a subcritical bifurcation takes place while approaching the codimension-2 line visualized by the solid line in Fig.~\ref{tricrit1}. Quite a similar behavior has been found in other systems, for instance in binary fluid convection \cite{Brand:1984,Cross:1988,Schöpf:1989}. 
%
%
If $D$ is in- or decreased compared to $D=0.5$ in Fig.~\ref{tricrit1}, the tricritical lines are moved along the corresponding codimension-2 lines towards more negative charge ratios while simultaneously abuting even further to the associated codimension-2 lines. Thus a broadening of the domain of supercritically bifurcating waves is observed for slightly negative charge ratios.
%
%
\subsubsection{Stability of traveling waves in absence of the large-scale field $C$}
\label{benjfeir}

Among the classes of solutions admitted by the complex Ginzburg--Landau equation Eq.~(\ref{ginzlandau}) is the family of plane wave solutions
\begin{align}\label{planewavesol}
A=\sqrt{\frac{\eta-q^2}{g_1}}\,e^{i(q\,x-\Omega\,t)}=Fe^{i(q\,x-\Omega\,t)}\,
\end{align}
with the frequency
\begin{align}
\Omega=-a-\eta c_1+(c_0-c_1)q^2
\end{align}
existing for $q^2<\eta$ in case of a supercritical bifurcation. One easily sees that solutions of the type (\ref{planewavesol}) are long-wavelength stable for a range of wave numbers surrounding the homogeneous $q=0$ state as long as the Benjamin--Feir criterion, $1+c_0c_1>0$, is satisfied \cite{Cross:1993,Aranson:2002,Feir:1967,Newell:1974,Matkowski:1993}.
%
\begin{figure}
\includegraphics [width=0.85\columnwidth] {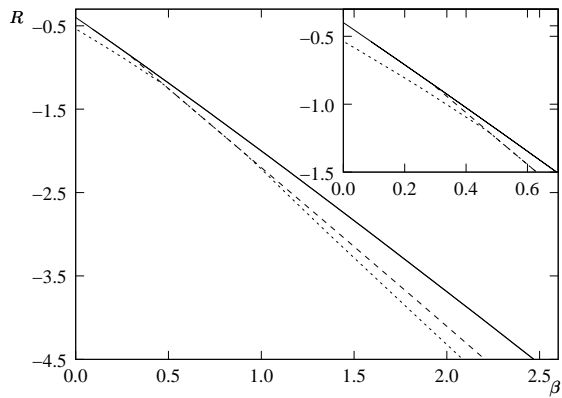}
\caption{\label{benjfeir1} The positions of the codimension-2 line (solid line), the tricritical line 
(dashed line) and the line caracterizing marginally long-wavelength stable plane waves (dotted line) 
are displayed in the $R$-$\beta$-plane. Traveling waves bifurcate supercritically in the area located 
underneath the tricritical line and plane wave solutions given by Eq.~(\ref{planewavesol}) are
long-wavelength unstable in the region lying in-between the tricritical line and the dotted line, 
wherein $c_0\,c_1<-1$ holds. Parameters used are $D=\beta_c=0.5$, as well as $\alpha=0.8$.
}
\end{figure}
%

Provided that the charge ratios $R$ are sufficiently negative, traveling waves are, according to Fig.~\ref{benjfeir1}, linearly stable in a large area of the $R$-$\beta$-plane underneath the dotted line, determined by the marginal stability criterion $1+c_1 c_0=0$, and become unstable in-between the latter line and the tricritical one visualized by the dashed line. Additionally a small unstable stripe, wherein the Benjamin--Feir criterion is violated, emerges for slightly negative charge ratios in the neighborhood of the codimension-2 line depicted by the solid line in Fig.~\ref{benjfeir1} wherein $D=\beta_c=0.5$ and $\alpha=0.8$.

\begin{figure}
\includegraphics [width=0.85\columnwidth] {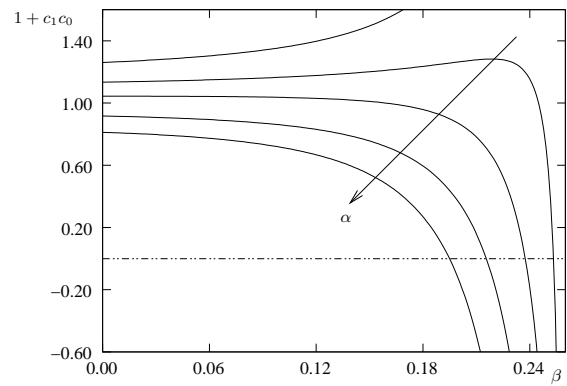}
\caption{\label{benjfeir2} The Benjamin--Feir criterion $1+c_1\,c_0=0$ for marginally long-wavelength 
stable plane wave solutions is plotted as a function of the conversion rate ratio $\beta$ while assuming
a slightly negative charge ratio $R=-0.8$. Along the directed arrow the density parameter $\alpha$ is
progressively increased from 0.2 to 0.8 by steps of 0.1 and evinces that for $\alpha\rightarrow 1$
the domain of unstable plane waves lying in the immediate neighborhood of the codimension-2 line is 
incessantly aggrandized. Remaining parameters read $D=\beta_c=0.5$.
}
\end{figure}

\begin{figure}
\includegraphics [width=0.85\columnwidth] {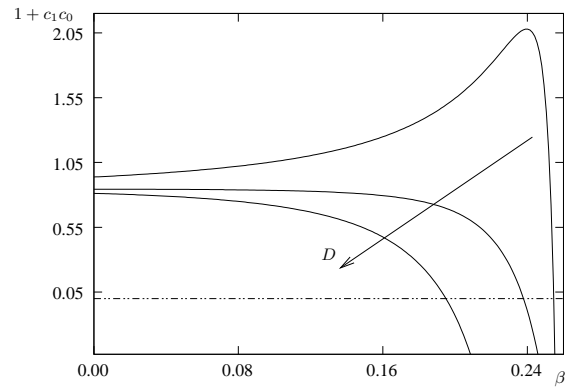}
\caption{\label{benjfeir3} The dependance of the marginal Benjamin--Feir criterion $1+c_1\,c_0=0$ on 
the diffusion ratio $D$ as a function of $\beta$ for a density parameter $\alpha=0.8$, the charge 
ratio $R$ evaluating to $-0.8$. Increasing the diffusion ratio from $D=0.3$ via $D=0.4$ to $D=0.5$ 
along the arrow leads to a maximal enlargement of the unstable regime of wave solutions, the latter 
effect being once again reversed if $D>0.5$. Furthermore $\beta_c=0.5$.
}
\end{figure}

Whereas the extension of the instability region at strongly negative charge ratios is barely affected by the choice of the density parameter $\alpha$, the one located at mildly negative values of $R$ utterly depends on $\alpha$ as shown exemplarily in Fig.~\ref{benjfeir2} for $R=-0.8$, the diffusion ratio $D$ as well as the conversion rate $\beta_c$ still evaluating to $0.5$. The marginal stability condition $1+c_1 c_0=0$ being plotted in Fig.~\ref{benjfeir2} as a function of the conversion ratio $\beta$ up to the associated codimension-2 point, it turns out that this unstable stripe vanishes for small density parameters. If $\alpha$ is in contrast kept constant at $0.8$ while the diffusion ratio $D$ is varied, Fig.~\ref{benjfeir3} proves that a diffusion ratio of $D=0.5$ induces a maximal broadening of this very unstable band situated at slightly negative charge ratios. For strongly negative charge ratios the region of Benjamin--Feir-stable traveling waves is widened in a similiar manner to the one of supercritically bifurcating waves presented in Fig.~\ref{tricrit2}.
%
%

The effect that traveling waves become unstable via Benjamin--Feir resonance, which may eventually lead to spatiotemporal chaos, while approaching the tricritical point lying in the vicinity of a codimension-2 point has been predicted in a general context \cite{Zimmermann:1988} and discussed within the scope of binary fluid convection \cite{Zimmermann:1993}.
%
%
\subsection{Stability of traveling waves in presence of the large-scale mode $C$}
\label{nonlinearwaves}
%
In the present section we consider once more traveling wave solutions of the form Eq.~(\ref{planewavesol}) and study, in contrast to Sec.~\ref{benjfeir}, their stability by implicating  the long-wavelength stationary excitation $C$, that is by taking into account the global conservation law imposed upon the ion channel densities. To analyse the stability of the aforesaid traveling waves we set
\begin{subequations}
\label{nonlinwavesol}
\begin{align}
A&=[F+v_1(x, t)]\,e^{\dot{\imath}(qx-\Omega t)}\, ,\\
C&=v_2(x ,t)\, ,
\end{align}
\end{subequations}
substitute these expressions into the amplitude equations Eqs.~(\ref{amplitudeeq}) and linearizing with respect to the small perturbation terms $v_1(x,t)$ and $v_2(x,t)$ yields in case of the homogeneous $q=0$ state the following coupled partial differential equations
\begin{subequations}
\label{nonlineqs}
\begin{align}
\tau[\partial_t +v_g\partial_x -\dot{\imath}\Omega]v_1
&=\left[(\eta+\dot{\imath} a)+\xi_0^2(1+\dot{\imath}c_0)\partial_x^2\right] v_1 \nonumber\\
&+F\left[g_5(1+\dot{\imath}c_5)\partial_x- g_2(1+\dot{\imath}c_2)\right]\,v_2\nonumber\\
&-g_1(1+\dot{\imath}c_1)(v_1^{\star}+2 v_1)F^2\, ,\\
\partial_t\,v_2 &=D_1\partial_x^2\,v_2 -b F\partial_x (v_1 + v_1^{\star})
\end{align}
\end{subequations}
with $F$ and $\Omega$ denoting amplitude respectively frequency of the plane wave solution Eq.~(\ref{planewavesol}). It is beneficial to consider sideband instability by setting
\begin{subequations}
\label{nonlinansatz}
\begin{align}
v_1(x,\, t)&=V_1\,e^{\Sigma t+\dot{\imath}K x}+V^{\star}_2\,e^{\Sigma^{\star} t-\dot{\imath}K x}\, ,\\
v_2(x, t)& =V_3\,e^{\Sigma t+\dot{\imath}K x}+V_4^{\star} e^{\Sigma^{\star} t-\dot{\imath}K x}\, ,
\end{align}
\end{subequations}
where $V_i\,(i=1,2,3,4)$ designate complex amplitudes. From the coupled equations Eqs.~(\ref{nonlineqs}) the aforementioned ansatz then generates a set of homogeneous, linear equations whose solubility condition leads to the growth rate $\Sigma(K)$ as a function of the perturbation wave number $K$. 

The expansion of the latter growth rate with respect to $K$
\begin{align}
\label{SigmaEnt}
\Sigma(K)\approx -\dot{\imath}\,v_g\,K - (1+c_0c_1)\,\frac{\xi_0^2}{\tau}\,K^2 + \mathcal{O}\,(K^4)
\end{align}
shows that in the limit of small perturbation wave numbers $K$ the Benjamin--Feir criterion of Sec.~\ref{benjfeir} remains always unchanged when taking into account the large-scale mode $C$ and switching to a frame moving with group velocity $v_g$. Since the large-scale mode $C$ leaves the stability of traveling waves untouched with respect to long-wavelength perturbations, it might alter their stability by triggering an instability at finite perturbation wave number which is assured by Fig.~\ref{nichtlinDispersio}$(b)$.
%
\begin{figure}
\includegraphics [width=0.85\columnwidth] {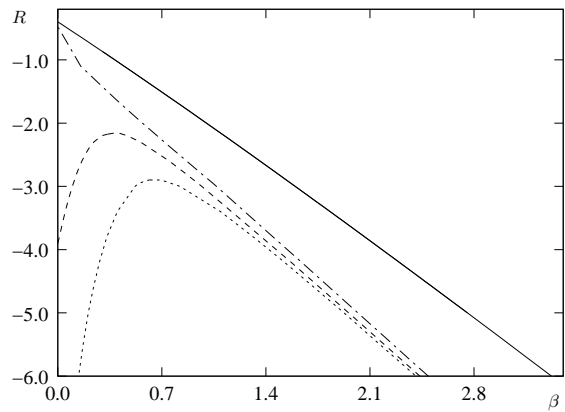}
\caption{\label{nonlinstab1} Stability regions of nonlinear traveling waves given by 
Eqs.~(\ref{nonlinwavesol}) within the $R$-$\beta$-plane for $D=\beta_c=0.5$ and different
values of the density parameter $\alpha$. The solid line visualizes the codimension-2 line,
traveling waves being stable in the area lying underneath the lines of marginal stability, 
namely the dashed-dotted one corresponding to $\alpha=0.6$, the dashed one to $\alpha=0.75$
and the dotted one to a density parameter of $\alpha=0.8$. The distance to the bifurcation 
point was chosen to be $\eta=0.03$.
}
\end{figure}
%

Compared to the case $C=0$ studied in Sec.~\ref{benjfeir} exemplarily for $D=\beta_c=0.5$ and $\alpha=0.8$ through Fig.~\ref{benjfeir1}, Fig.~\ref{nonlinstab1} clearly evinces a strong influence of the large-scale mode $C$ due to the conservation law on the stability of traveling waves: Regions of the  $R$-$\beta$-plane, wherein stable plane wave solutions occur, are significantly restricted and shifted towards more negative charge ratios. Fig.~\ref{nonlinstab1} shows those stability domains limited, depending on the choice of the density parameter $\alpha$, to the area located below the dashed-dotted, the dashed or the dotted line of marginally stable nonlinear traveling waves satisfying $\Sigma(K)=0$. The real part of the latter growth rate $\Sigma(K)$ is plotted in Fig.~\ref{nichtlinDispersio}$(b)$ as a function of the perturbation wave number $K$ presuming the density parameter $\alpha$ evaluates to 0.8. In account with the stability diagram, i.e. Fig.~\ref{nonlinstab1}, the tuple ($R=-5.5$, $\beta=2.17$) corresponds to wave solutions becoming unstable with respect to finite perturbation wave numbers $K$ as expected from the power series of $\Sigma(K)$ performed in Eq.~(\ref{SigmaEnt}). To hightlight the effect of the large-scale mode $C$ evoked by the conservation law, the real part $\Re(\Sigma)$ of the same growth rate is depicted in Fig.~\ref{nichtlinDispersio}$(a)$ in case of a vanishing real mode $C$. The stability analysis of $C=0$ in Sec.~\ref{benjfeir} has shown by means of the stability diagram Fig.~\ref{benjfeir1} that for the parameter set ($R=-4$, $\beta=1.86$, $\alpha=0.8$) traveling waves are long-wavelength unstable and thus the observed instability in Fig.~\ref{nichtlinDispersio}$(a)$ is generically different from the one encountered in the presence of a conserved quantity whose typical shape is given by Fig.\ref{nichtlinDispersio}$(b)$.
%
\begin{figure}
\begin{center}
\includegraphics [width=0.98\columnwidth] {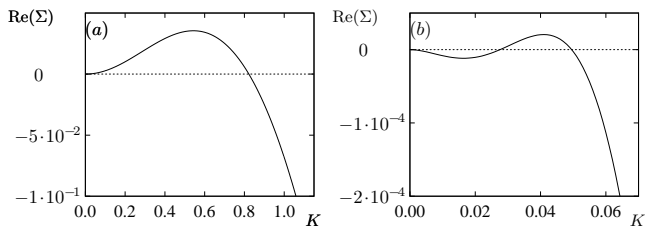}
\end{center}
\caption{\label{nichtlinDispersio} The typical shape of the real part $\Re(\Sigma)$ corresponding to the growth rate $\Sigma(K)$ specified by the ansatz Eq.~(\ref{nonlinansatz}) is plotted as a function of the perturbation wave number $K$ for $D=\beta_c=0.5$, $\alpha=0.8$ as well as $\eta=0.03$ in $(b)$. $R=-5.5$ and $\beta=2.17$ leads, according to the stability diagram in Fig.~\ref{nonlinstab1}, to wave solutions unstable with respect to
finite perturbation wave numbers $K$. On the left hand side however the tuple $(R=-4,\beta=1.86)$ induces, in the absence of the large-scale field $C$ characterizing the conservation law, generic long-wavelength unstable traveling waves in accordance with Fig.~\ref{benjfeir1}.
}
\end{figure}
%

\begin{figure}
\includegraphics [width=0.85\columnwidth] {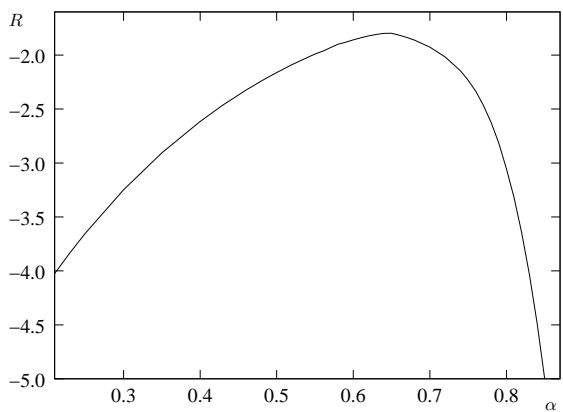}
\caption{\label{nonlinstab2} Points of marginal stability for a fixed conversion ratio $\beta=0.5$
as a function of the density parameter $\alpha$, the diffusion ratio $D$ and the conversion
rate $\beta_c$ evaluating to $0.5$ as in Fig.~\ref{nonlinstab1}. Furthermore $\eta=0.03$.
}
\end{figure}

Moreover the sensitivity of the former unstable stripe at slightly negative values of $R$ (cf. enlargement in Fig.~\ref{benjfeir1}) on the density parameter, depicted in Fig.~\ref{benjfeir2}, is amplified as it may be concluded from Fig.~\ref{nonlinstab1} as well as from Fig.~\ref{nonlinstab2}, wherein the points of marginally stable long-wavelength nonlinear plane wave solutions are plotted as a function of $\alpha$ for a constant conversion ratio $\beta$. The latter figure also stipulates that, for a given value of $\beta$, there exists a certain density parameter $\alpha$ generating a maximal outspread of the area of stable traveling waves. For the given parameter set $D=\beta_c=\beta=0.5$ and $\eta=0.03$ a density parameter of $\alpha\approx 0.65$, meaning the ion channels contribute up to $65\%$ to the total conductance of the membrane layer, realizes the largest extension of the stable traveling wave regime.
%
%
%
\subsection{Numerical results}
\label{numerics}
\begin{figure}
\begin{center}
\includegraphics [width=0.98\columnwidth]{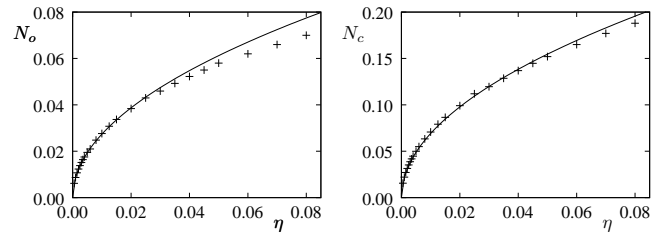}
\end{center}
\caption{\label{simulstripe} The amplitudes $N_o$ and $N_c$ of traveling 
(opened and closed respectively)
ion channel densities obtained by numerical simulations (crosses) of the
microscopic model given by Eqs.~(\ref{scaleq}) are compared
to the analytical solution 
$A =\sqrt{\eta/g_1}\,\evec{e}_0\,(\tilde N_o,\tilde N_c)^T$. The parameters
used are $D=1,\beta=0.4,R=-1.8,\alpha=0.8,\tau_v=0.01$ with the individual
conversion rates being $\beta_c=0.5$ and $\beta_o=0.2$. 
}
\end{figure}

%
A major problem always encountered when dealing with amplitude equations is that their validity range around the threshold $\eta=0$ remains a priori unpredictable. Therefore we have determined the amplitude of a stripe solution
by numerical simulations of the basic Eqs.~(\ref{scaleq}) as a function of the relative distance $\eta$ to the bifurcation point and compared these numerical results with the analytical solution $A=\sqrt{\eta/g_1}\,\evec{e}_0\,(\tilde N_o,\tilde N_c)^T$ in Fig.~\ref{simulstripe}. Up to $\eta \sim 0.05$ we find for the particle density $N_c$ of the closed ion channels a fairly good agreement between analytical and numerical results whereas the same holds up to $\eta \sim 0.03$ for the opened ion channel density $N_o$, which gives a reasonable estimate of the quantitative validity range of the preceding results 
obtained in terms  of amplitude equations.
%
%
\section{Discussion and conclusion}
\label{conclusion}

Besides its biological motivation, the model presented in this work exhibits interesting generic properties concerning pattern formation that are for the first time discussed within the scope of a specific model system. 
The homogeneous ion channel density distribution on the biomembrane may become unstable with respect to either a stationary or an oscillatory bifurcation, whereas we focused exclusively on analyzing the latter bifurcation type. As the total ion channel density is a conserved field, the supercritical Hopf-bifurcation occuring at finite wave number shows nonlinear properties largely differing from those emerging in systems lacking conserved order parameters.

In presence of a global conservation law, such as the one encountered in our model, it can be expected that the asymptotic description of pattern formation by means of amplitude equations is altered through the coupling of known forms of envelope equations to long-wavelength modes. This allegation has been confirmed by the perturbation expansion of the initial model equations that yielded a set of coupled amplitude equations in case of traveling ion channel density waves rather than a single complex Ginzburg-Landau equation. The validity range of the latter reduced dynamics has been checked by simulations of the inital model equations. This coupling to stationary, long-wavelength excitations has however been completely neglected during the discussion of the stationary bifurcation scenario in Ref.~\cite{KramerSC:2002} which gives consequently a distorted picture of the nonlinear bifurcation behavior.

Detailed phase diagrams highlight in which parameter space Hopf-bifurcations at finite wavelengths take place. Neither the threshold value of the control parameter nor the supercriticality of the oscillatory bifurcation are modified by implicating the long-wavelength mode evoked by the conserved ion channel density field into the asymptotic description near onset of instability. The stability of traveling ion channel density waves is however strongly affected by the latter real mode for infinitely extended systems. It turns out that the long-wavelength mode, reflecting the conserved density field, triggers an instability at finite perturbation wave number whereas plane wave solutions would be long-wavelength unstable in the absence of the large-scale mode. If compared to a case without a real mode, stable traveling wave solutions are generally shifted towards strongly negative ratios of opened and closed ion channel charges respectively and thus the domain, wherein spatiotemporal chaos is likely to be seen, significantly broadens. If finally all model parameters are fixed the observation of stable traveling ion channel density waves becomes less probable the more the ion channels contribute to the total conductance of the considered membrane layer.

The dynamics of traveling waves in finite geometries and consequently their stability properties naturally differ from those in infinitely extended systems \cite{Cross:1986.1,Cross:1988.1,Steinberg:1987,Ahlers:1987}. How the finite size of the system further influences the studied traveling ion channel density waves at hand, already affected by a large-scale mode, is postponed to future work.

The formation of dissipative patterns of ion channels on a membrane that were discussed in the present work is most likely to be observed in in vitro systems as the one proposed in Ref.~\cite{Fromherz:1998} where a linear array of 48 field-effect transistors in silicon was coated with a bimolecular layer of lipid and gramicidin. Nevertheless the investigated electrodiffusive mechanism seems to be also relevant for in vivo systems as shown by experiments on the effect of electric fields on clustering of acetylcholine receptors as cited in Ref. \cite{Leonetti:2005}.

Couplings between a long-wavelength mode and a traveling wave are expected in other models from cell biology as well, e.g. in mixtures of cytoskeletal filaments interacting with molecular motors \cite{Ziebert:2006}.
The investigation of an oscillatory instability occuring at finite wave number coupling to a stationary, long-wavelength mode, that was motivated by the biological model at hand, will be extended within the scope of a reaction-diffusion system in a forthcoming work to oscillatory, long-wavelength excitations influencing on oscillatory instabilities at finite wave number \cite{Peter:2006}.

\begin{center}
{\bf Acknowledgments}
\end{center}
Fruitful discussions with M. Hilt and F. Ziebert were very much
appreciated.

%
%
\appendix
\section{Amplitude expansion}
\label{appA}
According to the slow time and spatial scales $T_1$, $T$ and $X$ introduced in Sec.~\ref{genericamplitueeq} all time and space derivatives occuring in the scaled equations Eqs.~\eqref{scaleq} may be substituted by
\begin{subequations}\label{slowvariables}
\begin{align}
\partial_t &\to \partial_t + \sqrt{\eta}\,\partial_{T_1} + \eta\,\partial_T\, ,\\
\partial_x  &\to \partial_x + \sqrt{\eta}\,\partial_{X}\, .
\end{align}
\end{subequations}
The pattern-generating ion channel densities $N_o$ and $N_c$ as well as the transmembrane voltage $V$ have to be expanded with respect to powers of $\sqrt{\eta}$
\begin{subequations}
\begin{align}
N_o &= \sqrt{\eta}\,N_o^1 +\eta\,N_o^2 +\eta^{\frac{3}{2}}\,N_o^3
                           +{\cal O}\left(\eta^2\right)\, ,\\
N_c &= \sqrt{\eta}\,N_c^1 +\eta\,N_c^2 +\eta^{\frac{3}{2}}\,N_c^3 
                           +{\cal O}\left(\eta^2\right)\, ,\\
V   &= \sqrt{\eta}\,V^1 +\eta\,V^2 +\eta^{\frac{3}{2}}\,V^3 +{\cal O}\left(\eta^2\right)
\end{align}
\end{subequations}
in order to accomplish a perturbation expansion of Eqs.~\eqref{scaleq}. Within the adiabatic limit, i.e. if instantaneously adapting membrane voltage $V$ is assumed, meaning formally $C_m=0$ or $\tau_v=0$, Eq.~\eqref{scaleq3} becomes time independant and may be solved successively by means of the resulting equation hierarchy in $V^i\quad(i=1,2,3)$  
\begin{subequations}
\begin{align}
\eta^{\frac{1}{2}}:\,\,\mathcal{F}\,V^1 &= -e_c N_o^1\, ,\\
\eta^{\frac{2}{2}}:\,\, \mathcal{F}\,V^2 &= -e_c N_o^2
                                              +(2\partial_x\partial_X-\alpha N_o^1)V^1\, ,\\
\eta^{\frac{3}{2}}:\,\,\mathcal{F}\,V^3 &= -e_c (N_o^1+N_o^3)+(\partial_X^2-\alpha N_o^2)V^1\nonumber\\&\hspace{0.5cm}+(2\partial_x\partial_X-\alpha N_o^1)V^2\, ,	      
\end{align}
\end{subequations}
yielding the three leading order contributions to $V$
\begin{subequations}
\label{Viter}
\begin{align}
V^1 & = \mathcal{F}^{-1}\left[-e_c^H N_o^1\right]\, ,\\
V^2 & = \mathcal{F}^{-1}\left[-e_c^H N_o^2 +(2\partial_x\partial_X -\alpha N_o^1) V^1\right]\, ,\\
V^3 &= \mathcal{F}^{-1}\left[-e_c^H (N_o^1+N_o^3) +(\partial_X^2 -\alpha N_o^2)V^1\right.\nonumber\\
    &\hspace{1.2cm}\left.+(2\partial_x\partial_X -\alpha N_o^1)V^2\right]
\end{align}
\end{subequations}
wherein the operator
\begin{align}
\mathcal{F}=1-\partial_x^2
\end{align}
has been introduced.\\
Since the desired solutions  $N_o^i$ and $N_c^i$ with $i=1,2,3$ are harmonic functions of
$e^{i(\omega_c t \pm k_c x)}$, the occuring operator  $\mathcal{F}^{-1}$ in the antecedent expressions for $V^i$ may be replaced in the forthcoming calculations as 
\begin{align}
\label{diffop}
\mathcal{F}^{-1}\,\, e^{n\dot{\imath}\left(k_c x +\omega_c t\right)}
\longmapsto \dfrac{1}{1+n^2 k_c^2}\, e^{n\dot{\imath}\left(k_c x +\omega_c t\right)}
\end{align}
with $n=0,1,2,\ldots$ 

By performing the perturbation expansion of the scaled equations Eqs.~\eqref{scaleq1},~\eqref{scaleq2} governing the spatiotemporal evolution of the ion channel densities and taking into account Eqs.~\eqref{Viter} as well as Eq.~\eqref{diffop} one ends up with a hierarchy in $\sqrt{\eta}$
\begin{subequations}
\label{hie}
\begin{align}
\eta^\frac{1}{2} :&\,\mat{L}_0 \vec{w}_1 =0\label{hie1}\\
\eta^\frac{2}{2} :&\,\mat{L}_0 \vec{w}_2 =\mat{N}_1\left(\vec{w}_1\right)
                           -\left(\mat{L}_1 +\mat{M}_0\partial_{T_1}\right)\vec{w}_1\label{hie2}\\
\eta^\frac{3}{2} :&\,\mat{L}_0 \vec{w}_3 =\mat{N}_2\left(\vec{w}_1 ,\vec{w}_2\right)
                 -\left(\mat{L}_1+\mat{M}_1\partial_{T_1}\right)\,\vec{w}_2\nonumber\\
 &\qquad\qquad\qquad\qquad\quad-\left(\mat{L}_2+\mat{M}_2\partial_{T}\right)\,\vec{w}_1 \label{hie3}
\end{align}
\end{subequations}
wherein the occuring differential operators read
\begin{align}
\vec{w}_i &=\left(\begin{array}{c} N_o^i \\ N_c^i \end{array}\right),\quad (i=1,2,3)\, ,\\
\mat{M}_0 &=\mat{M}_1 =\mat{M}_2=\left(\begin{array}{cc} 1 & 0\\ 0 & 1\end{array}\right)\, ,
\end{align}
\begin{widetext}
\begin{subequations}
\begin{align}
\mat{L}_0 =&\renewcommand{\arraystretch}{2}\left(\begin{array}{cc}
                   \partial_t-\partial_x^2 +e_c^H\partial_x^2\,\mathcal{F}^{-1} +\beta_c 
		   & -\beta_c\\
                   DQe_c^H\partial_x^2\,\mathcal{F}^{-1} -\beta_o & 
		   \partial_t-D\partial_x^2 +\beta_o
                   \end{array}
             \right)\, ,\qquad
\mat{L}_1 =2\renewcommand{\arraystretch}{2}\left(\begin{array}{cc}
                 -1+e_c^H \mathcal{F}^{-2} & 0\\
                  DQe_c^H \mathcal{F}^{-2} & -D
                  \end{array}
             \right)\partial_x\partial_X\, ,\\
\mat{L}_2 =&\renewcommand{\arraystretch}{2}\left(\begin{array}{cc}
              -\partial_X^2 +e_c^H\left[\left(\partial_x^2
	       +\partial_X^2\right)\,\mathcal{F}^2+5\partial_x^2\partial_X^2\,\mathcal{F}
        +4\partial_x^4\partial_X^2\right]\mathcal{F}^{-3} & 0\\
              DQe_c^H\left[\left(\partial_x^2
	       +\partial_X^2\right)\,\mathcal{F}^2+5\partial_x^2\partial_X^2\,\mathcal{F}
        +4\partial_x^4\partial_X^2\right]\mathcal{F}^{-3} & -D\partial_X^2
                \end{array}\right)\, ,\\
\mat{N}_1 =& -\left(\begin{array}{c} 1\\DQ \end{array}\right)
             \left[-\mathcal{F}^{-1}\partial_x^2\left(N_o^1\,\mathcal{F}^{-1}
	     N_o^1\right)\right]\alpha e_c^H
           \renewcommand{\arraystretch}{2}-\left(\begin{array}{c}
              \partial_x N_o^1\,\mathcal{F}^{-1} N_o^1
              +N_o^1\,\mathcal{F}^{-1}\partial_x^2 N_o^1
              \\ DQ\left[\partial_x N_c^1\,\mathcal{F}^{-1} N_o^1
                +N_c^1\,\mathcal{F}^{-1}\partial_x^2 N_o^1\right]
                \end{array}\right)e_c^H\, ,\\
\mat{N}_2 =& -\left(\begin{array}{c} 1\\DQ \end{array}\right)
               \left[-\mathcal{F}^{-1}\partial_x^2\left(N_o^2\,\mathcal{F}^{-1}
	       N_o^1\right)
               -2\mathcal{F}^{-1}\partial_x\partial_X(1+\partial_x^2)\left(N_o^1\,\mathcal{F}^{-1} N_o^1\right)
               -m_1\right]\nonumber\\
           &\qquad-e_c^H\renewcommand{\arraystretch}{2}\left(\begin{array}{c}
                     m_2 \partial_x N_o^1 +\left(\partial_X N_o^1+\partial_x N_o^2
                     +N_o^2\partial_x\right)\mathcal{F}^{-1}\partial_x N_o^1 +m_3
		      N_o^1\\
                     DQ\left[m_2\partial_x N_c^1 +\left(\partial_X N_c^1+\partial_x N_c^2
                     +N_c^2\partial_x\right)\mathcal{F}^{-1}\partial_x N_o^1+m_3 
		     N_c^1\right]
                     \end{array}\right)
\end{align}
with the abbreviations
\begin{align}
m_1 =& -\partial_x\mathcal{F}^{-1}
         \left[N_o^1\left(\mathcal{F}^{-1}N_o^2+2\partial_x\partial_X\mathcal{F}^{-2}N_o^1
               -\alpha\mathcal{F}^{-1}\left(N_o^1\mathcal{F}^{-1}N_o^1\right)
	       \right)\right]\, ,\\
m_2 =& \left(2\partial_x\partial_X\mathcal{F}^{-2}+\partial_X\mathcal{F}^{-1}\right)N_o^1
       +\mathcal{F}^{-1}N_o^2\, ,\\
m_3 =& \left(2\partial_x^3\partial_X\mathcal{F}^{-2}+2\partial_x\partial_X\mathcal{F}^{-1} \right)N_o^1 +\partial_x^2\mathcal{F}^{-1}N_o^2\, .
\end{align}
\end{subequations}
\end{widetext}
The aforementioned hierarchy, Eqs.~\eqref{hie}, has to be solved successively for $\vec{w}_i$ with $i=1,2,3$. The first of these three equations is fulfilled if the ansatz given in Eq.~\eqref{amplinlo} is chosen for $\vec{w}_1$. The determined solution may be inserted into equation Eq.~\eqref{hie2} which leads to a linear, inhomogeneous equation of the type $\mat{L}_0\vec{w}_2=\mat{I}_2$. Since $\mat{L}_0$ is not invertible, as it has a zero eigenvalue, the only possibility for the latter equation to have nontrivial solutions is that $\mat{I_2}$ is orthogonal to the kernel of $\mat{L}_0^\dagger$ which yields the solvability condition 
\begin{align}
&\langle\vec{f}_0\,e^{\dot{\imath}({\bf k_c^H}\cdot\vec r+\omega_c t)}|\mat{I}_2\rangle\label{fred1}\\
&\,\,=\frac{1}{S}\,\int_S\,d\vec r\,\frac{1}{\overline T}\,\int_{\overline T}\,dt\,e^{-\dot{\imath}({\bf k_c^H}\cdot\vec r+\omega_c t)}\,\vec{f}_0^\dagger\,\mat{I}_2=0
\nonumber
\end{align}
with $\vec{f}_0$ denoting the left eigenvector of $\mat{L}_0$. Terms proportional to $e^0$, linked to the large-scale mode $C$, stipulate a second Fredholm's alternative
\begin{align}\label{fred2}
\langle\vec{z}_0|\mat{I}_2\rangle=0
\end{align}
where $\vec{z}_0=\left(1,-\beta\right)^\top$ designates the left eigenvector spanning the adjungated kernel of $\left.\mat{L}_0\right|_{k=0}$. Inserting the solution $\vec{w}_2$ of Eq.~\eqref{hie2} as well as $\vec{w}_1$ into Eq.~\eqref{hie3} gives rise to another linear, inhomogeneous equation of the form $\mat{L}_0\vec{w}_3=\mat{I}_3$. To the latter equation are associated two solvability equations
\begin{align}
\langle\vec{f}_0\,e^{\dot{\imath}({\bf k_c^H}\cdot\vec r+\omega_c t)}|\mat{I}_3\rangle=0\, ,\label{fred3}\\
\langle\vec{z}_0|\mat{I}_3\rangle=0\, .\label{fred4}
\end{align}
From Fredholm's alternatives at order $\eta$, namley Eqs.~\eqref{fred1} and \eqref{fred2}, as well as at order $\sqrt{\eta}^3$, that is Eqs.~\eqref{fred3} and \eqref{fred4}, the final amplitude equations \eqref{amplitudeeq} can be obtained after rescaling space and time according to
\begin{align}
\partial_t A=\sqrt{\eta}\,\partial_{T_1} A + \eta\,\partial_{T} A\, ,\\
\partial_t C=\sqrt{\eta}\,\partial_{T_1} C + \eta\,\partial_{T} C
\end{align}
as suggested by \eqref{slowvariables}.


%
%
\bibliographystyle {prsty}

\end{document}